\documentclass{elsart3p}

 \usepackage{epsfig}

\usepackage{amssymb}

\begin{document}

\begin{frontmatter}

\title{Properties of the $\pi$ state induced by impurities in a d-wave 
superconductor}
 
\author[a]{Grzegorz Litak}
\author[b]{Mariusz Krawiec}
\address[a]{Department of Mechanics, Technical University of Lublin,
Nadbystrzycka 36, PL-20-618 Lublin, Poland}
\address[b]{Institute of Physics and Nanotechnology Center, M. Curie-Sk\l{}odowska 
University,
Radziszewskiego 10, \\ PL-20-031 Lublin, Poland}

\begin{abstract}
We study the properties of  a quantum impurity embedded
 in a superconducting
 host of a d-wave symmetry. The superconductor is described by the extended 
negative 
$U$
 Hubbard model while
the impurity introduces a repulsive interaction to the system around the central 
site.
We discuss
 the influence of this repulsion on the local
 properties
 (density of states, electron pairing and spontaneous current around
 the
 impurity)  of the superconductor. We show the condition of $\pi$ - like
 behaviour, defined
 as two subsystems having  a phase difference  of $\pi$, in the
 system by using a proper
 combination of attractive pairing bond interaction and
 repulsive one between electrons located at the impurity and nearest neighbour sites.
\end{abstract}

\end{frontmatter}


Impurities in superconductors ($SC$) has been a subject of
intensive  theoretical and experimental studies
\cite{Balatsky2006,Tallon1997,Karpinska2000}
recently.
In case of a nonuniform d-wave superconductor 
(surfaces, vortices, cracks, twin boundaries or impurities)
$\pi$-phase jump in the $SC$ order parameter can be seen 
\cite{pi_junction_Ka,pi_junction_Lo}.
In this case Josephson current becomes
negative
in
contradiction to the usual $0$-phase junction. The situation is similar in 
granular
high-$T_c$ materials which can likely form network of microscopic
$\pi$-junctions \cite{Sigrist1995} between small regions with different phases of
the order parameter. In such systems the zero-energy Andreev bound states,
zero-bias conductance peaks, paramagnetic Meisner effect and spontaneously
generated currents take place \cite{pi_junction_Ka,pi_junction_Lo}.
The situation is also  possible for paramagnetic impurities with  
strong on-site Coulomb repulsion are placed in classical $s$-wave
superconductor \cite{Litak2005,Litak2006,Lambert1998}. 
Due to the proximity effect  the $SC$ order
parameter is created on the impurity site but change its phase by $\pi$.

The purpose of the present work to check if 
an analogous solution is possible for an impurity with  a repulsive interaction 
embedded in a $d$-wave 
paired superconductor.

Let us describe the system  by the extended negative $U$ Hubbard model 
\cite{Micnas1991} 
with
the
Hamiltonian:

\begin{eqnarray}
 H = \sum_{ij\sigma} \left(t_{ij} - \mu \delta_{ij}\right)
     c^+_{i\sigma} c_{j\sigma} +
     \frac{1}{2} \sum_{ij\sigma} U_{ij} n_{i\sigma} n_{j-\sigma},
 \label{Hamiltonian}
\end{eqnarray}
where $i$, $j$ label sites of a square lattice, $t_{ij} = -t$ is the hopping
integral between nearest neighbour sites and $\mu$ - chemical potential.
$U_{ij} < 0$ describes attraction between electrons with opposite spins occupying
neighbour sites $i$ and $j$. The effect of impurity is introduced in one of the 
lattice
sites $i=0$ via repulsive interaction $U_{0j} > 0$.

In the following we shall work in the Hartree-Fock approximation dropping the
Hartree terms, which means that we have only
anomalous part impurity (in a paring potential $\Delta_{ij}$) induced disorder in 
the system. So the corresponding
Gorkov equation has the form:

\begin{eqnarray}
 \sum_{j'} \left( 
 \begin{array}{c}
  (\omega + \mu)\delta_{ij'} - t_{ij'};~ \Delta_{ij'}  \\
  \Delta^{\ast}_{ij'};~  (\omega + \mu)\delta_{ij'} + t_{ij'}
 \end{array}  
 \right)
 \hat G(j',j;\omega) = \delta_{ij}.
 \label{Gorkov}
\end{eqnarray}
In the zero temperature $SC$ order parameter $\Delta_{ij}$ and the total local
charge $n_i$ are given by self-consistent relations:
%
\begin{eqnarray}
&& \Delta_{ij} \equiv U_{ij} \chi_{ij} =
            -  U_{ij} \frac{1}{\pi}\int^{E_f}_{-\infty} {\rm d}\omega \;
            {\rm Im} G^{12}(i,j;\omega),
\label{charge}  \\
&&  n_i = -2 \int^{E_f}_{-\infty} {\rm d}E D(E) =-  
\frac{2}{\pi}\int^{E_f}_{-\infty} {\rm d}\omega \;
            {\rm Im} G^{11}(i,i;\omega),
\nonumber
\end{eqnarray} 
where $E_f$ denotes Fermi energy and 
$D(E)$ defines the quasiparticle density of 
states.
\begin{figure}[tbp] 
\epsfig{file=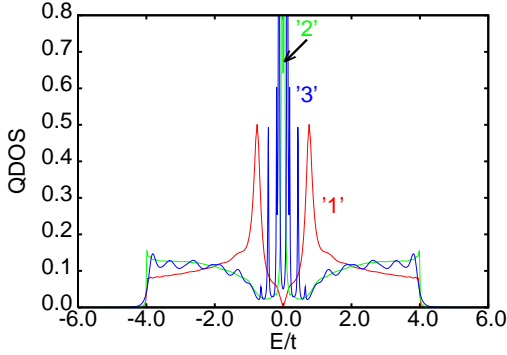,width=5.5cm,angle=-90}
\caption{Quasiparticle density of states $D(E)$ for a pure superconductor '1',
$D(E)$ at the impurity site without electric current around the 
impurity '2' and with 
current  '3'.}
\end{figure}
Eqs. (\ref{Gorkov}-\ref{charge}) have been solved  self-consistently
at the two dimensional lattice using a recursion method 
\cite{recursion1}.
Looking for
the pairing
amplitude 
at the impurity site
$\Delta_{0j}$, 
we assumed for simplicity that the surrounding
sites may be treated homogeneously as in bulk (clean) superconcuctor.
Calculations were performed for interaction parameters: attractive for a bulk
$U_{ij}/t = -0.975$, repulsive for an impurity interactions $U_{0j}/t =3.0$
and a half filled system
($n=1$).
In Fig. 1 the curve '1'
corresponds to the quasi-particle density of states for a pure superconductor '1' 
while
'2' to the local density of states at the impurity site
with the $\pi$ phase change in $\Delta_{0j}$.
The density of states represented by the curve '2' shows two very close peaks located 
symmetrically near Fermi energy ($E=0$) signaling bound states within the 
energy gap and the appearance of the $\pi$ state. 
Applying an external magnetic field perpendicular to the surface, one can 
shift those states away the zero energy (curve 3 in Fig. 1). The shifting is 
proportional to vector potential ${\bf A}$ included for a region close to 
the impurity through a Peierls substitution \cite{Peierls1933,Miller1995} 
\begin{equation}
t_{ij} \rightarrow t_{ij} \exp{\frac{-{\rm i}e}{\hbar } \int_{\bf r_i}^{\bf r_j} 
{\bf A} ( {\bf r}') {\rm d} {\bf r}'}.
\end{equation}
The applied magnetic field generates a current flowing around the impurity.

Those preliminary results will be further checked by a self-consistent 
calculations with respect to paring amplitude $\Delta_{ij}$, current and the 
vector potential  in order to see if a spontaneous current will be generated in 
such a system \cite{Krawiec2002}.

{\bf Acknowledgements:} This work has been partially supported 
by KBN grant No. 2P03B06225. GL would like to thank Max Planck Institute for the 
Physics of Complex Systems in Dresden for hospitality.


\begin{thebibliography}{00}
\bibitem{Balatsky2006}
A.V. Balatsky {\it et
al.}, 
Rev. Mod. Phys. 78, (2006) 373.
\bibitem{Tallon1997}
J.L. Tallon {\it et
al.}, Phys. Rev. Lett. {\bf 79} (1997) 5294.
\bibitem{Karpinska2000}
 K. Karpi\'nska
{\it et al.},
Phys. Rev. Lett. 84 (2000) 610.
\bibitem{pi_junction_Ka} S. Kashiwaya and Y. Tanaka, Rep. Prog. Phys.  
63 (2000) 1641.
\bibitem{pi_junction_Lo}
                      T. L\"{o}fwander {\it et al.}, Superconduct. Sci.   
Technol. 14, (2001) R53.
\bibitem{Sigrist1995} M. Sigrist and T. M. Rice, Rev. Mod. Phys. 67
                  (1995) 503.
\bibitem{Litak2005}  G. Litak and M. Krawiec, Phys. Stat. Sol. B 242 (2005) 
438.
\bibitem{Litak2006}  G. Litak and  M. Krawiec, Physica B 378-380 (2006), 434.


\bibitem{Lambert1998} C.J. Lambert and R. Raimondi, J.  Phys. Condens. Matter 
                  10 (1998) 901.

\bibitem{Micnas1991} R. Micnas {\it et al.}, Rev. Mod. Phys. 62 (1991) 113.

\bibitem{recursion1} G. Litak {\it et al.}, Physica C 251 (1995) 263.
 

\bibitem{Peierls1933} R.E. Peierls, Z. Physik 80 (1933) 763.
\bibitem{Miller1995} P. Miller, B.L. Gyorffy, J. Phys. Condens. Matter 7
(1995) 5579. 

\bibitem{Krawiec2002} M. Krawiec {\it et al.},
Phys. Rev. B 66 (2002) 172505.

\end{thebibliography}
\end{document}